\documentclass[conference, onecolumn]{IEEEtran}
\IEEEoverridecommandlockouts
\usepackage{cite}
\usepackage{amsmath,amssymb,amsfonts}
\usepackage{algorithmic}
\usepackage{graphicx}
\usepackage{textcomp}
\usepackage{xcolor}
\def\BibTeX{{\rm B\kern-.05em{\sc i\kern-.025em b}\kern-.08em
    T\kern-.1667em\lower.7ex\hbox{E}\kern-.125emX}}
\begin{document}

\title{Secrecy Limits of Energy Harvesting IoT Networks under Channel Imperfections}

  \author{\IEEEauthorblockN{Furqan Jameel\IEEEauthorrefmark{1}, Zheng Chang\IEEEauthorrefmark{2}, Riku J\"antti\IEEEauthorrefmark{1}}
                                         
                                          \IEEEauthorblockA{\IEEEauthorrefmark{1} Department of Communications and Networking, Aalto University, 02150 Espoo, Finland. \\ \IEEEauthorrefmark{2} Faculty of Information Technology, University of Jyv\"askyl\"a, Finland.}
}

\maketitle

\begin{abstract}
Simultaneous wireless information and power transfer (SWIPT) has recently gathered much research interest from both academia and industry as a key enabler of energy harvesting Internet-of-things (IoT) networks. Due to a number of growing use cases of such networks, it is important to study their performance limits from the perspective of physical layer security (PLS). With this intent, this work aims to provide a novel analysis of the ergodic secrecy capacity of a SWIPT system is provided for Rician and Nakagami-$m$ faded communication links. For a realistic evaluation of the system, the imperfections of channel estimations for different receiver designs of the SWIPT-based IoT systems have been taken into account. Subsequently, the closed-form expressions of the ergodic secrecy capacities for the considered scenario are provided and, then, validated through extensive simulations. The results indicate that an error ceiling appears due to imperfect channel estimation at high values of signal-to-noise ratio (SNR). More importantly, the secrecy capacity under different channel conditions stops increasing beyond a certain limit, despite an increase of the main link SNR. The in-depth analysis of secrecy-energy trade-off has also been performed and a comparison has been provided for imperfect and perfect channel estimation cases. As part of the continuous evolution of IoT networks, the results provided in this work can help in identifying the secrecy limits of IoT networks in the presence of multiple eavesdroppers.
\end{abstract}

\begin{IEEEkeywords}
IoT Network, Eavesdroppers; Nakagami-m fading; Physical layer security; Rician fading; SWIPT
\end{IEEEkeywords}

\maketitle

\section{Introduction}
Internet-of-things (IoT) is gradually maturing due to a considerable amount of research interest received in recent years \cite{hassija2019survey}. It seems evident that the myriad applications of IoT in healthcare, industrial automation, and agriculture, are going to usher an era of massive connectivity of millions of devices. According to the researchers, the increase in IT consolidation and the rapid adoption of reliable solutions are some of the main contributors to such massive connectivity of wireless IoT devices \cite{hou2019survey}. Modernization of legacy infrastructure is another focal point to provide seamlessly connected solutions that have a significant impact on the performance of such massive IoT networks. In this regard, ensuring the security of the IoT devices is at the forefront of critical issues. Due to this reason, a great number of research efforts are directed towards securing IoT devices against a number of attacks \cite{8437135}. Another important aspect of such networks is energy consumption and reliability on rechargeable battery solutions. In particular, there is a growing demand to develop energy harvesting solutions to reduce the dependency and consumption of fossil fuels and move towards green communications \cite{8214104}. To ensure such ubiquitous connectivity of green IoT networks, several interesting yet daunting issues need to be addressed for future IoT networks.

The first challenge stems in the form of prevention against eavesdropping attacks. Due to broadcast nature, it seems inevitable that the wireless links among IoT devices would be susceptible to the eavesdropping attacks. Moreover, since many of the IoT networks may operate privately in ad-hoc mode, the security techniques involving a trusted third party may not be realizable. In this regard, one of the promising solutions has been proposed in the form of physical layer security (PLS). The basic principle of PLS is to ensure link security by harnessing the randomness of the wireless channel \cite{jameel2017secrecy2}. This randomness of the wireless channel can be used to confuse the eavesdropper and reliably transfer the message to the intended receiver. Moreover, it can ensure good communication secrecy without requiring exhaustive cryptographic operations \cite{jameel2017secrecy}. Besides this, message authentication can also be performed for ensuring the confidentiality of messages in the presence of malicious IoT devices. 

\begin{figure}
    \centering
        \begin{tabular}{c}
        \includegraphics[width=.23\textwidth]{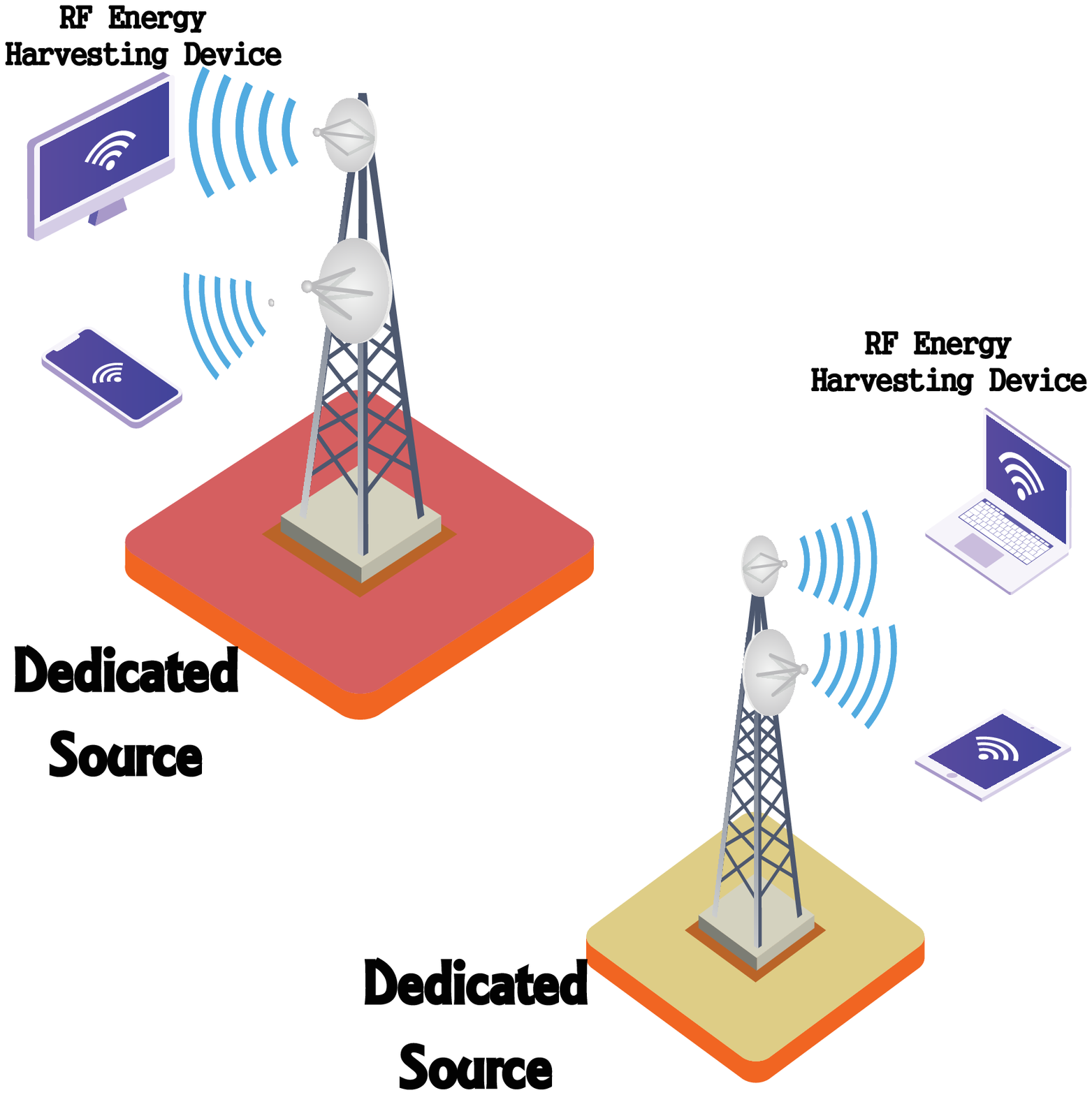}\\
(a) \\
        \includegraphics[width=.23\textwidth]{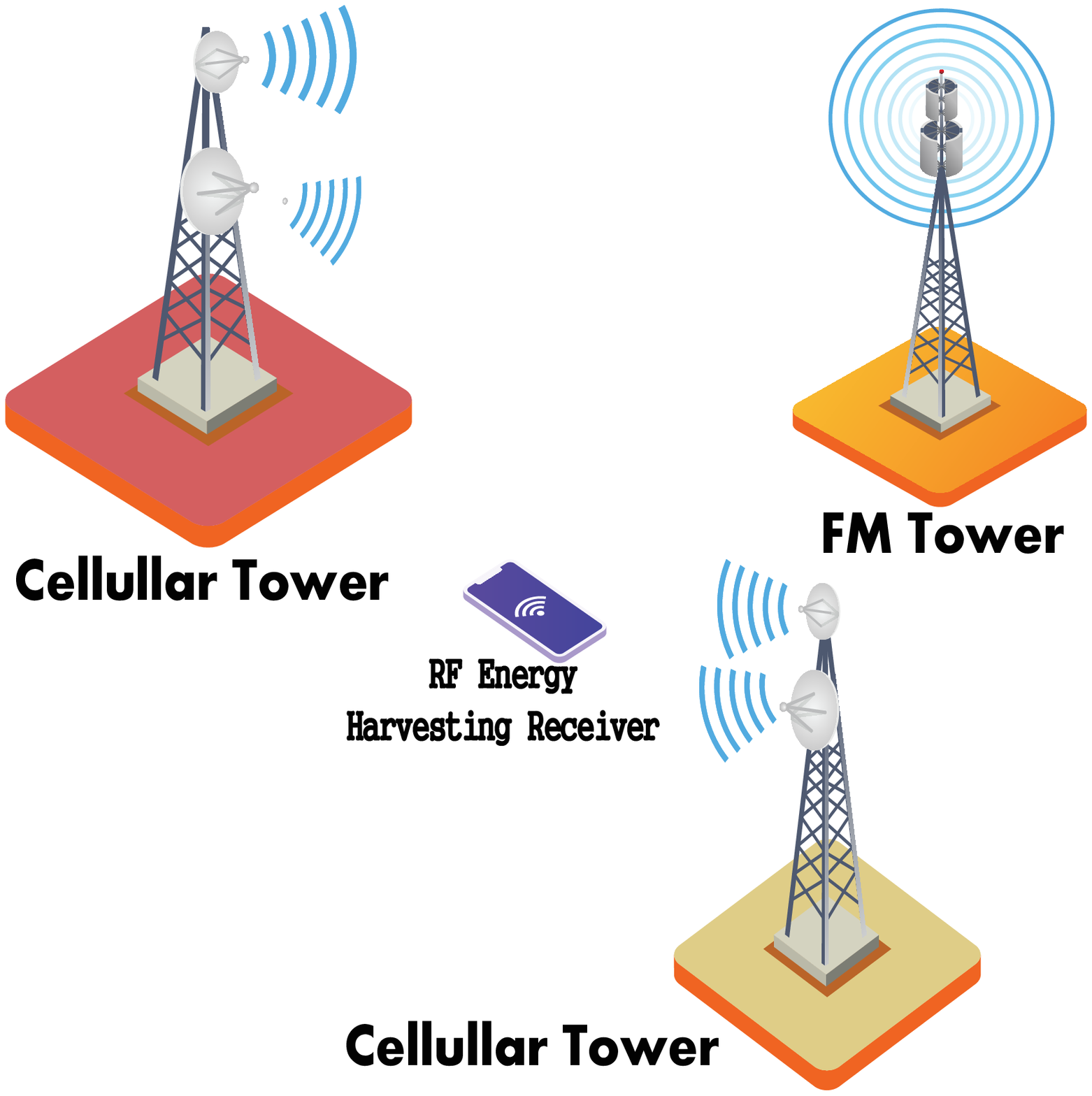}\\
(b) \\
    \end{tabular}
        \caption{RF energy harvesting architectures (a) dedicated energy harvesting (b) ambient energy harvesting.}
    \label{fig:block}
\end{figure}

The second issue is related to the limited life-cycle of wireless devices which can be addressed by employing efficient energy harvesting techniques \cite{krikidis2014simultaneous}. The term energy harvesting which is sometimes called power harvesting is a technique that allows devices to harvest energy from already available sources. For RF energy harvesting, these sources include radio, cellular and television broadcasting signals that are mostly available and free to use for harvesting. This idea also gave birth to the transfer of power and data through electromagnetic (EM) waves \cite{8246039}. As shown in Figure \ref{fig:block}, the transfer of energy can take place using dedicated or ambient resources in the environment. The dedicated energy harvesting can be performed using dedicated towers to transfer energy to remote devices, whereas, ambient energy harvesting can be performed using radio frequency (RF) waves generated from wireless devices and fixed nodes. This dual nature of EM waves (i.e., ability to transfer information and energy) has given birth to a promising field of research namely simultaneous wireless information and power transfer (SWIPT) \cite{krikidis2014simultaneous}. 

As mentioned earlier, the inherent nature of wireless IoT devices makes them vulnerable to eavesdropping attacks \cite{nguyen2019secrecy}. In this regard, it is worth noting increasing the power for ensuring the reliability of messages can have a negative impact on the secrecy performance of the network. In case, when a power receiver is a malicious IoT device, then improving the power transfer efficiency can reduce secrecy. It is because by increasing the power of the information signal, one may reduce the secrecy rate of the wireless links \cite{ko2019neighbor}. In this context, few studies have recently explored the concept of PLS and energy harvesting to efficiently use the transmit power. For instance, the authors of \cite{liu2013secrecy} proposed a band allocation scheme, whereby, the band reserved for energy harvesting cannot be used for information decoding. In another study, the concept of secrecy design for low-powered devices was proposed and evaluated \cite{8377127}. They introduced power splitting (PS) and time switching (TS) techniques for information processing and energy harvesting. Security in full-duplex (FD) relays with SWIPT systems was studied in \cite{taghizadeh2016physical}. An optimization problem was formulated for minimizing power consumption and ensuring link security for multiple-input-multiple-output (MIMO) FD relays. In another work \cite{bi2016accumulate}, Bi \emph{et al}. proposed a strategy based on energy accumulation and then jamming the eavesdropper reception. The authors of \cite{wu2017secure} provided a secrecy maximization scheme for energy harvesting relays. They first formulated a utility maximization problem and then solved it with the help of the Lagrange dual method.

The main contribution of this paper lies in the performance evaluation of energy harvesting and secure IoT. Specifically, we consider two fading channel models i.e., Rician and Nakagami-m fading. We derive closed-form expressions of ergodic secrecy capacity for both separated and integrated SWIPT-enabled IoT device architecture. In contrast to the conventional approach, we consider imperfect channel estimation for both legitimate and eavesdropping links. To validate our findings, we perform Monte-Carlo simulations in MATLAB.

The rest of the paper is arranged as follows. Section 2 provides details of the considered system model. In Section 3, a detailed analysis of ergodic secrecy capacity for both Nakagami-m and Rician fading has been provided. Section 4 provides analytical and simulation results along with their relevant discussion. Lastly, Section 5 presents some key conclusions. 

\section{System Model}

Let us consider a downlink communication network constituting an access point (AP) and $N+1$ SWIPT-enabled IoT devices. Within any timeslot, the AP communicates with a legitimate IoT device ($\mathcal{S}$) while the other $N$ IoT devices can be considered as eavesdroppers denoted as $\mathcal{E}=\{E_i|i=1,2,...N\}$. All the IoT devices are equipped with single antennas having the capability to harvest energy and decode information from the received signal. In order to efficiently use the transmitted power from AP, during each time slot, the AP transmits information to $\mathcal{S}$ while the other devices are supposed to harvest energy from the radio waves. Being part of the same network, the energy harvesting IoT devices may act as eavesdroppers. Furthermore, in order to decode information and harvest energy $\mathcal{S}$ and $\mathcal{E}$ are assumed to use PS scheme \cite{huang2013simultaneous}. As per PS policy, the signal at the receiving IoT device is divided into two streams for energy harvesting and information decoding. Particularly, the multiplicating PS factor $\rho$ is used for information decoding and $1-\rho$ portion of the power is used for harvesting energy. 

\subsection{Fading Channel Models}
 
For many tropospheric, urban, sub-urban and indoor communication systems, the channel is an amalgamation of noise, fading, and interference. Therefore, an appropriate selection of the channel fading model is important to analyze the performance of a wireless network. In this context, we adopt two renowned channel fading models namely, Rician and Nakagami-$m$ distribution, for characterization of secrecy performance of SWIPT-enabled IoT devices. Rice distribution has been used to model environments where a strong stable line of sight component exists between source and destination. The theoretical model is the sum of two vectors, one with Rayleigh distributed modulus and other with known modulus with a large value. The Nakagami-$m$ distribution is a multi-faceted distribution which is used to model numerous fading environments. It has proven to be highly flexible for matching empirical data of indoor and outdoor environments. It includes Rayleigh and One-sided Gaussian as special cases. Some studies also indicate that Nakagami-$m$ can be used to approximate Rician fading for its shape parameter $m>>1$ \cite{nakagami1960m}. 

\subsection{Signal Propagation}

When AP, with power $P$, transmits its signal $s$ to $\mathcal{S}$, the received signal is written as \cite{zhou2013wireless}
\begin{align}
y_{s}=\sqrt{\rho 
_{s}}\left(\sqrt{\frac{P}{P_{s}^{loss}}}\hat{h}_{s}s+n_{s}\right)+z_{s},
\end{align}
where $\hat{h}_s$ is the wireless channel gain between $\mathcal{S}$ and AP, and $\rho_s$ is the PS factor for $\mathcal{S}$. The signal processing noise is denoted as $z_s$, $n_s$ is the additive white Gaussian noise (AWGN) given as $\mathcal{N}(0,\sigma_s^2)$ and $P_s^{loss}$ is the path loss. As explained earlier, the communication is overheard by the eavesdropping IoT devices. At the $i$-th eavesdropper, the received signal is expressed as 
\begin{align}
y_{ie}=\sqrt{\rho 
_{ie}}\left(\sqrt{\frac{P}{P_{ie}^{loss}}}\hat{h}_{ie}s+n_{ie}\right)+z_{ie},
\label{hat}
\end{align}
where $\hat{h}_{ie}$ is the channel gain and $P_{ie}^{loss}$ is the path loss, $z_{ie}$ represents the signal processing noise, and $n_{ie}$ denotes the AWGN between the AP and $i$-th eavesdropper. We assume that the eavesdropper use similar hardware and $\sigma_{ie}^2=\sigma_{e}^2$ $\forall \ E$. For tractability of analysis, we consider $n_{ie}=n_e$ and $\rho_{ie}=\rho_e$ and $P_{ie}^{loss}=P_e^{loss}$, $\forall \ i \ \in \ N$. In this paper, we consider imperfect channel estimation which is more practical as it is generally difficult to perfectly estimate wireless channel. Thus, a renowned channel estimation model has been used to incorporate channel imperfections in our system \cite{isukapalli2010packet,yoo2006capacity} 
\begin{align}
\hat{h}_{k}=\sqrt{1-\delta _{k}^{2}}h_{k}+\delta _{k}v,
\end{align}
where $h_k$ is the estimated wireless channel gain and $k\in (s,ie)$. Also,  $\delta_k$ denotes the channel accuracy factor wherein for $\delta_k = 0$ the channel is assumed to be accurately known and  for $\delta_k = 1$ the channel estimation is completely inaccurate. Moreover, $v$ is given by $\mathcal{N}(0,1)$. Since we are focusing on the secrecy performance evaluation of the system, we assume that the value of $\delta_k$ is given a priori. By substituting above equation in (\ref{hat}) we obtain
\begin{align}
y_{s}&=\sqrt{\rho _{s}}\left(\sqrt{\frac{P(1-\delta 
_{s}^{2})}{P_{s}^{loss}}}h_{s}s+\sqrt{\frac{P}{P_{s}^{loss}}}\delta 
_{s}vs+n_{s}\right)+z_{s}, \\
y_{ie}&=\sqrt{\rho _{ie}}\left(\sqrt{\frac{P(1-\delta 
_{ie}^{2})}{P_{e}^{loss}}}h_{ie}s+\sqrt{\frac{P}{P_{e}^{loss}}}\delta 
_{ie}vs+n_{ie}\right) \nonumber  \\
&+z_{ie}.
\end{align}

Using above equations, the instantaneous received SNR $\mathcal{S}$ can be expressed as 
\begin{align}
\gamma _{s}=\frac{\rho _{s}\Omega _{s}(1-\delta _{s}^{2})}{(\Omega 
_{s}\rho _{s}\delta _{s}^{2}+\rho _{s}N_{0}+\sigma _{s}^{2})}\vert 
h_{s}\vert ^{2},
\label{eqr_1}
\end{align}

Similarly, the instantaneous received SNR at any eavesdropper is written as
\begin{align}
\gamma _{ie}=\frac{\rho _{e}\Omega _{e}(1-\delta _{ie}^{2})}{(\Omega 
_{e}\rho _{s}\delta _{ie}^{2}+\rho _{e}N_{0}+\sigma _{e}^{2})}\vert 
h_{ie}\vert ^{2},
\label{eqr_2}
\end{align}
where $\Omega_s=P/(P_s^{loss})$ and $\Omega_e=P/(P_e^{loss})$ are the average transmit power of main and wiretap link in (\ref{eqr_1}) and (\ref{eqr_2}), respectively. It is worth mentioning that among all the eavesdroppers, an eavesdropper with the best CSI has a better chance to decode the received message. Thus, the received SNR of eavesdropping link yields
\begin{align}
\gamma _{e}=\max _{i \in N}\gamma _{ie}=\frac{\rho _{e}\Omega 
_{e}(1-\delta _{e}^{2})}{(\Omega _{e}\rho _{s}\delta _{e}^{2}+\rho 
_{e}N_{0}+\sigma _{e}^{2})}\max _{i \in N} \vert h_{ie}\vert ^{2}.
\end{align}

\section{Ergodic Secrecy Capacity}
An important security indicator is the average secrecy capacity $C_{sec}$. It is expressed as
\begin{align}
\bar{C}_{sec}=\mathbf{E}\lbrace C_{sec}\rbrace =\mathbf{E}\lbrace \lbrack 
C_{s}-C_{e}\rbrack ^{+}\rbrace .
\label{av_c_sec}
\end{align}

Now, we first discuss both channels separately.

\subsection{Ergodic Secrecy Capacity under Rician Fading}

\subsubsection{Separated $\mathcal{S}$ and $\mathcal{E}$}

By using (\ref{av_c_sec}) we get
\begin{align}
\bar{C}_{sec }&=\mathbf{E}\lbrace [\log_{2}(1+\gamma _{s})-\log _{2}(1+\gamma 
_{e})]^{+}\rbrace \nonumber \\
&=\int_{0}^{\infty }{\underbrace{\int_{0}^{\gamma _{s}}{\lbrack \log 
_{2}(1+u)-\log (1+v)\rbrack }f_{\gamma _{e}}(v)dv}_{\mathcal{G}_{1}}} \nonumber \\
&\times f_{\gamma _{s}}(u)du.
\label{der_a_s_sec}
\end{align}

Now by using integration by parts, we get

\begin{align}
\mathcal{G}_{1}=\frac{1}{\ln 2}\int_{0}^{u}{\frac{F_{\gamma _{e}}(v)}{1+v}dv}.
\label{g_1}
\end{align}

Replacing (\ref{g_1}) in (\ref{der_a_s_sec}) and changing the order of integration, we obtain
\begin{align}
\bar{C}_{sec}&=\frac{1}{\ln 2}\int_{0}^{\infty }{\frac{F_{\gamma 
_{e}}(v)}{1+v}\left[ \int_{v}^{\infty }{f_{\gamma _{s}}(v)}du\right] 
dv} \nonumber \\
&=\frac{1}{\ln 2}\int_{0}^{\infty }{\frac{F_{\gamma 
_{e}}(v)}{1+v}\lbrack 1-F_{\gamma_{s}}(v)\rbrack dv}.
\label{a_c_sec}
\end{align}

Using \cite{zogas2005infinite}, the probability density function (PDF) of the legitimate link is written as 
\begin{align}
f_{\gamma _{s}}(\gamma _{s})&=\frac{(\Omega _{s}\rho _{s}\delta 
_{s}^{2}+\rho _{s}N_{0}+\sigma _{s}^{2})(K_{s}+1)}{\rho _{s}(1-\delta 
_{s}^{2})\bar{\gamma }_{s}}\nonumber \\
& \times \exp\left(-\frac{(\Omega _{s}\rho _{s}\delta 
_{s}^{2}+\rho _{s}N_{0}+\sigma _{s}^{2})(K_{s}+1)\gamma _{s}}{\rho 
_{s}(1-\delta _{s}^{2})\bar{\gamma }_{s}}\right) \nonumber \\
& \times I_{0}\left(2\sqrt{\frac{(\Omega _{s}\rho _{s}\delta _{s}^{2}+\rho 
_{s}N_{0}+\sigma _{s}^{2})K_{s}(K_{s}+1)\gamma _{s}}{\rho _{s}(1-\delta 
_{s}^{2})\bar{\gamma }_{s}}}\right) \nonumber \\
& \times \exp\left(-K_{s}\right). 
\label{Rice_pdf_s}
\end{align}

Similarly, the cumulative distribution function (CDF) of the received SNR at $\mathcal{S}$ is given as 

\begin{align}
&F_{\gamma _{s}}(\gamma _{s})= \nonumber \\
&1-Q_{1}\left(\sqrt{2K_{s}},\sqrt{\frac{2(\Omega _{s}\rho _{s}\delta 
_{s}^{2}+\rho _{s}N_{0}+\sigma _{s}^{2})(K_{s}+1)\gamma _{s}}{\rho 
_{s}(1-\delta _{s}^{2})\bar{\gamma }_{s}}}\right) ,
\label{Rice_cdf_s}
\end{align}
where $\bar{\gamma}_s=\Omega_s \mathbf{E}\{|h_s|^2\}$ denotes the mean value, $I_0 (.)$ is the modified Bessel function of the first kind and order zero \cite{gradshteyn2014table} and $Q$ is the Marcum-$Q$ function \cite{gradshteyn2014table} which is given as 
\begin{align}
Q_{1}(a,b)=\int_{b}^{\infty}{x} \exp(-\frac{(x^{2}+a^{2})}{2})I_{0}(ax)dx.
\end{align} 

Additionally, $K_s$ is the ratio of the main link's power of the line of sight component to the scattered component.

By substituting (\ref{Rice_cdf_s}) in (\ref{a_c_sec}), we get (\ref{eq_121}) at the top of next page. Now with the help of \cite[(1.211.1),(3.194.3)]{gradshteyn2014table}, ergodic secrecy capacity can be obtained as

\begin{figure*}
\vspace{-0.5cm}
\begin{align}
\bar{C}_{sec}^{Sp-Sp}&=\frac{1}{\ln 2}\sum_{z=0}^{N}{\biggl(\begin{matrix}
\ N & \\
\ z & \\
\end{matrix}
\biggr)(-1)^{z}}\int_{0}^{\infty }\frac{1}{1+\gamma _{e}} \times \exp \left[ -z\exp(\upsilon 
(\sqrt{2K_{e}}))\left\{ \frac{2(K_{e}+1)(\Omega _{e}\rho _{e}\delta 
_{e}^{2}+\rho _{e}N_{0}+\sigma _{e}^{2})\gamma _{e}}{\rho 
_{e}\bar{\gamma }_{e}(1-\delta _{e}^{2})}\right\}^{\frac{\mu 
(\sqrt{2K_{e}})}{2}} \right] \nonumber \\
& \times {\exp \left[-\exp(\upsilon (\sqrt{2K_{s}}))\left\{ \frac{2(K_{s}+1)(\Omega _{s}\rho _{s}\delta _{s}^{2}+\rho 
_{s}N_{0}+\sigma _{s}^{2})\gamma _{e}}{\rho _{s}(1-\delta 
_{s}^{2})\bar{\gamma }_{s}}\right\} ^{\frac{\mu 
(\sqrt{2K_{s}})}{2}}\right]}d\gamma _{e}.
\label{eq_121}
\end{align}
\hrulefill
\end{figure*}
\begingroup\makeatletter\def\f@size{7}\check@mathfonts
\def\maketag@@@#1{\hbox{\m@th\large\normalfont#1}}
\begin{align}
&\bar{C}_{sec}^{Sp-Sp}=\frac{1}{\ln 2}\sum_{z=0}^{N}{\biggl(\begin{matrix}
\ N & \\
\ z & \\
\end{matrix}}
\biggr)(-1)^{z} \mathbf{B}\left(\frac{\mu (\sqrt{2K_{e}})}{2},-\frac{\mu (\sqrt{2K_{e}})}{2}\right) \nonumber \\
&\times \left[ 1-z\exp(\upsilon (\sqrt{2K_{e}}))\left\{ 
\frac{2(K_{e}+1)(\Omega _{e}\rho _{e}\delta _{e}^{2}+\rho 
_{e}N_{0}+\sigma _{e}^{2})}{\rho _{e}\bar{\gamma }_{e}(1-\delta 
_{e}^{2})}\right\} ^{\frac{\mu (\sqrt{2K_{e}})}{2}} \right] \nonumber \\
& -\frac{1}{\ln 2}\sum_{z=0}^{N}{\biggl(\begin{matrix}
\ N & \\
\ z & \\
\end{matrix}}
\biggr)(-1)^{z} \times \mathbf{B}\left(\frac{\mu (\sqrt{2K_{s}})}{2},-\frac{\mu 
(\sqrt{2K_{s}})}{2}\right) \nonumber \\
& \times \exp(\upsilon (\sqrt{2K_{s}}))\left[ \frac{2(K_{s}+1)(\Omega _{s}\rho _{s}\delta 
_{s}^{2}+\rho _{s}N_{0}+\sigma _{s}^{2})}{\rho _{s}(1-\delta 
_{s}^{2})\bar{\gamma }_{s}}\right] ^{\frac{\mu 
(\sqrt{2K_{s}})}{2}}. 
\label{Rice_sp_sp} 
 \end{align}\endgroup

where $\mathbf{B}(.,.)$ is the incomplete Beta function \cite{gradshteyn2014table}. 

\subsubsection{Separated $\mathcal{S}$ and Integrated $\mathcal{E}$}

Using the method presented above sub-section, we can easily derive the expression of ergodic secrecy capacity when $\mathcal{S}$ and $\mathcal{E}$ employ separated and integrated receivers, respectively. It is given by 

\begin{align}
\bar{C}_{sec}=\frac{1}{\ln 2}\int_{0}^{\infty }{\frac{F_{\gamma 
_{e}}(v)}{Cv}\lbrack 1-F_{\gamma_{s}}(v)\rbrack dv}.
\label{a_c_sec2}
\end{align}

Replacing (\ref{Rice_cdf_s}) in (\ref{a_c_sec2}) yields (\ref{eq_122}) at the top of next page. Using \cite[(3.194.3)]{gradshteyn2014table} along with Marcum $Q$ approximation in (\ref{eq_122}), we get

\begin{figure*}
\vspace{-0.5cm}
\begin{align}
\bar{C}_{sec }^{Sp-In}&=\frac{1}{\ln 2} \times \sum_{z=0}^{N}{\biggl(\begin{matrix}
N & \\
z & \\
\end{matrix}
\biggr)(-1)^{z}} \times \int_{0}^{\infty }\frac{1}{C\gamma _{e}} \times \exp \left[ -z\exp(\upsilon (\sqrt{2K_{e}}))\left[ \frac{2(K_{e}+1)(\Omega _{e}\rho _{e}\delta 
_{e}^{2}+\rho _{e}N_{0}+\sigma _{e}^{2})\gamma _{e}}{\rho 
_{e}\bar{\gamma }_{e}(1-\delta _{e}^{2})}\right] ^{\frac{\mu 
(\sqrt{2K_{e}})}{2}} \right] \nonumber \\
& \times {\exp \left[-\exp(\upsilon (\sqrt{2K_{s}})) \left[ 
\frac{2(K_{s}+1)(\Omega _{s}\rho _{s}\delta _{s}^{2}+\rho 
_{s}N_{0}+\sigma _{s}^{2})\gamma _{e}}{\rho _{s}(1-\delta 
_{s}^{2})\bar{\gamma }_{s}}\right] ^{\frac{\mu 
(\sqrt{2K_{s}})}{2}}\right]}d\gamma _{e}.
\label{eq_122}
\end{align}
\hrulefill
\end{figure*}

\begingroup\makeatletter\def\f@size{7}\check@mathfonts
\def\maketag@@@#1{\hbox{\m@th\large\normalfont#1}}
\begin{align}
&\bar{C}_{sec}^{Sp-In}=\frac{1}{C\ln 2}\sum_{z=0}^{N}{\biggl(\begin{matrix}
\ N& \\
\ z& \\
\end{matrix}}
\biggr)(-1)^{z}\times \left[ \mathbf{B}\left(\frac{\mu (\sqrt{2K_{e}})}{2},-\frac{\mu (\sqrt{2K_{e}})}{2}\right) \right. \nonumber \\
& \left.\vphantom{2} \times \left\{1-z\exp(\upsilon (\sqrt{2K_{e}}))\left( 
2(K_{e}+1)\frac{(\Omega _{e}\rho _{e}\delta _{e}^{2}+\rho 
_{e}N_{0}+\sigma _{e}^{2})}{\rho _{e}\bar{\gamma }_{e}(1-\delta 
_{e}^{2})}\right) ^{\frac{\mu (\sqrt{2K_{e}})}{2}} \right\} \right. \nonumber \\
& \left.\vphantom{2} -\exp(\upsilon (\sqrt{2K_{s}}))\left( 2(K_{s}+1)\frac{(\Omega _{s}\rho _{s}\delta 
_{s}^{2}+\rho _{s}N_{0}+\sigma _{s}^{2})}{\rho _{s}(1-\delta 
_{s}^{2})\bar{\gamma }_{s}}\right) ^{\frac{\mu 
(\sqrt{2K_{s}})}{2}} \right. \nonumber \\
& \left.\vphantom{2} \times \mathbf{B}\left(\frac{\mu (\sqrt{2K_{s}})}{2},-\frac{\mu 
(\sqrt{2K_{s}})}{2}\right) \right] . 
\label{Rice_sp_in}
\end{align}
\endgroup

\subsection{Ergodic Secrecy Capacity under Nakagami-$m$ Fading}

The PDF of the main link is written as \cite{karagiannidis2003multivariate}
\begin{align}
f_{\gamma _{s}}(\gamma _{s})&=\left[ \frac{m_{s}(\Omega _{s}\rho _{s}\delta _{s}^{2}+\rho 
_{s}N_{0}+\sigma _{s}^{2})}{\rho _{s}(1-\delta _{s}^{2})\bar{\gamma 
}_{s}}\right] ^{m_{s}} \nonumber \\
& \times\frac{(\gamma _{s})^{m_{s}-1}\exp\left(-\frac{m_{s}(\Omega _{s}\rho _{s}\delta 
_{s}^{2}+\rho _{s}N_{0}+\sigma _{s}^{2})\gamma _{s}}{\rho _{s}(1-\delta 
_{s}^{2})\bar{\gamma }_{s}}\right)}{\Gamma(m_{s})},
\label{Nak_pdf_s}
\end{align} 

Similarly, the CDF of received SNR at $\mathcal{S}$ is expressed as

\begingroup\makeatletter\def\f@size{9}\check@mathfonts
\def\maketag@@@#1{\hbox{\m@th\large\normalfont#1}}
\begin{align}
F_{\gamma _{s}}(\gamma _{s})&=1-\exp\left(-\frac{m_{s}(\Omega _{s}\rho 
_{s}\delta _{s}^{2}+\rho _{s}N_{0}+\sigma _{s}^{2})\gamma _{s}}{\rho 
_{s}(1-\delta _{s}^{2})\bar{\gamma}_{s}}\right)\sum_{r=0}^{m_{s}-1}\frac{1}{r!} \nonumber \\
&\times \left[ \frac{m_{s}(\Omega_{s}\rho _{s}\delta _{s}^{2}+\rho _{s}N_{0}+\sigma _{s}^{2})\gamma 
_{s}}{\rho _{s}(1-\delta _{s}^{2})\bar{\gamma }_{s}}\right]^{r}.
\label{Nak_cdf_s}
\end{align} \endgroup

where $m_s$ represents the total number of multipath clusters for main link. Using the derivation for CDF and PDF of above section, we get CDF as
\begingroup\makeatletter\def\f@size{9}\check@mathfonts
\def\maketag@@@#1{\hbox{\m@th\large\normalfont#1}}
\begin{align}
F_{\gamma _{e}}(\gamma _{e})&=\left[ 1-\exp\left(-\frac{m_{e}(\Omega _{e}\rho 
_{e}\delta _{e}^{2}+\rho _{e}N_{0}+\sigma _{e}^{2})\gamma _{e}}{\rho 
_{e}(1-\delta _{e}^{2})\bar{\gamma}_{e}}\right)\sum_{r=0}^{m_{e}-1}\frac{1}{r!} \right. \nonumber \\
&\left.\vphantom{2} \times \left[ \frac{m_{e}(\Omega_{e}\rho _{e}\delta _{e}^{2}+\rho _{e}N_{0}+\sigma _{e}^{2})\gamma 
_{e}}{\rho _{e}(1-\delta _{e}^{2})\bar{\gamma }_{e}}\right]^{r} \right]^{N}, 
\label{Nak_cdf_e}
\end{align} \endgroup

In a similar way, the PDF of $\gamma_e$ is obtained as

\begin{align}
f_{\gamma _{e}}(\gamma _{e})&=\frac{N(\gamma _{e})^{m_{e}-1}}{\Gamma 
(m_{e})}\left[\frac{m_{e}(\Omega _{e}\rho _{e}\delta _{e}^{2}+\rho 
_{e}N_{0}+\sigma _{e}^{2})}{\rho _{e}(1-\delta _{e}^{2})\bar{\gamma 
}_{e}}\right]^{m_{e}} \nonumber \\
& \times \left[ 1-\exp\left(-\frac{m_{e}(\Omega_{e}\rho _{e}\delta _{e}^{2}+\rho _{e}N_{0}+\sigma _{e}^{2})\gamma _{e}}{\rho _{e}(1-\delta _{e}^{2})\bar{\gamma}_{e}}\right) \right. \nonumber \\
& \left.\vphantom{2}  \times \sum_{r=0}^{m_{e}-1}\frac{1}{r!} \times \left\{ \frac{m_{e}(\Omega_{e}\rho _{e}\delta _{e}^{2}+\rho _{e}N_{0}+\sigma _{e}^{2})\gamma_{e}}{\rho _{e}(1-\delta _{e}^{2})\bar{\gamma }_{e}}\right\}^{r}\right] ^{N-1}\nonumber \\
& \times\exp \left(-\frac{m_{e}(\Omega _{e}\rho _{e}\delta 
_{e}^{2}+\rho _{e}N_{0}+\sigma _{e}^{2})\gamma _{e}}{\rho _{e}(1-\delta 
_{e}^{2})\bar{\gamma }_{e}}\right). 
\label{Nak_pdf_e}
\end{align}
where $m_e$ represents the total number of multipath clusters for wiretap link.

\subsubsection{Separated $\mathcal{S}$ and $\mathcal{E}$}

By replacing (\ref{Nak_cdf_s}) and (\ref{Nak_cdf_e}) in (\ref{a_c_sec}), we obtain
\begin{align}
&\bar{C}_{sec}^{Sp-SP}=\frac{1}{\ln 2}\int_{0}^{\infty 
}{\frac{F_{\gamma _{e}}(\gamma _{e})}{1+\gamma _{e}}\lbrack 1-F_{\gamma 
_{s}}(\gamma _{e})\rbrack d\gamma _{e}} \nonumber \\
& =\frac{1}{\ln 2}\int_{0}^{\infty }{\frac{1}{1+\gamma _{e}}}\left[ 1-\exp\left(-\frac{(\Omega _{e}\rho _{e}\delta _{e}^{2}+\rho _{e}N_{0}+\sigma_{e}^{2})\gamma _{e}}{\rho _{e}\bar{\gamma }_{e}(1-\delta _{e}^{2})}\right) \right. \nonumber \\
& \left. \vphantom{2} \times \sum_{r=0}^{m_{e}-1}\frac{1}{r!}\left\{ \frac{(\Omega 
_{e}\rho _{e}\delta _{e}^{2}+\rho _{e}N_{0}+\sigma _{e}^{2})\gamma 
_{e}}{\rho _{e}\bar{\gamma }_{e}(1-\delta _{e}^{2})}\right\} ^{r}\right]^{N} \nonumber \\
& \times \Gamma \left(m_{s},\frac{m_{s}(\Omega _{s}\rho _{s}\delta _{s}^{2}+\rho 
_{s}N_{0}+\sigma _{s}^{2})\gamma _{e}}{\rho _{s}(1-\delta 
_{s}^{2})\bar{\gamma}_{s}}\right)d\gamma _{e}.
\end{align}

Using \cite[(8.352.4)]{gradshteyn2014table} we obtain
\begingroup\makeatletter\def\f@size{7}\check@mathfonts
\def\maketag@@@#1{\hbox{\m@th\large\normalfont#1}}
\begin{align}
&\bar{C}_{sec }^{Sp-SP}=\frac{1}{\ln 2}\sum_{z=0}^{N}{\biggl(\begin{matrix}
\ N& \\
\ z& \\
\end{matrix}
\biggr)\frac{(-1)^{z}}{\Gamma (m_{s})\Gamma 
(m_{e})}}\sum_{a=0}^{m_{e}-1}\nonumber \\
& \times {\sum_{b=0}^{m_{s}-1}{\frac{(m_{e}-1)!(m_{s}-1)!}{(a!)^{z}b! 
}}} \int_{0}^{\infty }\frac{1}{1+\gamma _{e}} \nonumber \\
& \times \exp\left(-a\frac{(\Omega 
_{e}\rho _{e}\delta _{e}^{2}+\rho _{e}N_{0}+\sigma _{e}^{2})\gamma 
_{e}}{\rho _{e}\bar{\gamma }_{e}(1-\delta _{e}^{2})}-\frac{m_{s}(\Omega 
_{s}\rho _{s}\delta _{s}^{2}+\rho _{s}N_{0}+\sigma _{s}^{2})\gamma 
_{e}}{\rho _{s}(1-\delta _{s}^{2})\bar{\gamma }_{s}}\right) \nonumber \\
& \times \frac{(\Omega 
_{e}\rho _{e}\delta _{e}^{2}+\rho _{e}N_{0}+\sigma _{e}^{2})}{\rho 
_{e}\bar{\gamma }_{e}(1-\delta _{e}^{2})}\frac{m_{s}(\Omega _{s}\rho 
_{s}\delta _{s}^{2}+\rho _{s}N_{0}+\sigma _{s}^{2})\gamma _{e}^{2}}{\rho 
_{s}(1-\delta _{s}^{2})\bar{\gamma }_{s}}d\gamma _{e}.
\label{Nak_sp_sp_cap}
\end{align}\endgroup

According to the author's best knowledge, the above integral cannot be further simplified. But it can be readily evaluated with MATLAB, Mathematica or any other standard computational software.

\subsubsection{Separated $\mathcal{S}$ and Integrated $\mathcal{E}$}

Substituting (\ref{Nak_cdf_s}) and (\ref{Nak_cdf_e}) in (\ref{a_c_sec2}) yields
\begingroup\makeatletter\def\f@size{7}\check@mathfonts
\def\maketag@@@#1{\hbox{\m@th\large\normalfont#1}}
\begin{align}
& \ \ \bar{C}_{sec }^{Sp-In}=\frac{1}{\ln 2}\int_{0}^{\infty 
}{\frac{F_{\gamma _{e}}(\gamma _{e})}{C\gamma _{e}}\lbrack 1-F_{\gamma 
_{s}}(\gamma _{e})\rbrack d\gamma _{e}}. \nonumber \\
&=\frac{1}{\ln 2}\sum_{z=0}^{N}{\biggl(\begin{matrix}
\ N & \\
\ z & \\
\end{matrix}
\biggr)\frac{(-1)^{z}}{\Gamma (m_{s})\Gamma 
(m_{e})}}\sum_{a=0}^{m_{e}-1}{\sum_{b=0}^{m_{s}-1}{\frac{(m_{e}-1)!(m_{s}-1)!}{C(a!)^{z}b! 
}}} \nonumber \\
& \times \int_{0}^{\infty } \exp \left(-a\frac{(\Omega _{e}\rho _{e}\delta 
_{e}^{2}+\rho _{e}N_{0}+\sigma _{e}^{2})\gamma _{e}}{\rho 
_{e}\bar{\gamma }_{e}(1-\delta _{e}^{2})}-\frac{m_{s}(\Omega _{s}\rho 
_{s}\delta _{s}^{2}+\rho _{s}N_{0}+\sigma _{s}^{2})\gamma _{e}}{\rho 
_{s}(1-\delta _{s}^{2})\bar{\gamma }_{s}}\right) \nonumber \\
&\times \frac{m_{e}m_{s}(\Omega 
_{e}\rho _{e}\delta _{e}^{2}+\rho _{e}N_{0}+\sigma _{e}^{2})(\Omega 
_{s}\rho _{s}\delta _{s}^{2}+\rho _{s}N_{0}+\sigma _{s}^{2})\gamma 
_{e}}{\rho _{e}\rho _{s}(1-\delta _{e}^{2})\bar{\gamma }_{e}(1-\delta 
_{s}^{2})\bar{\gamma }_{s}}d\gamma _{e}.
\end{align}\endgroup

Now with the help of \cite[(3.326.2)]{gradshteyn2014table}, we get 
\begin{align}
&\bar{C}_{sec }^{Sp-In}=\frac{1}{\ln 2}\sum_{z=0}^{N}{\biggl(\begin{matrix}
\ N & \\
\ z & \\
\end{matrix}
\biggr)\frac{(-1)^{z}}{\Gamma (m_{s})\Gamma 
(m_{e})}}\sum_{a=0}^{m_{e}-1}{\sum_{b=0}^{m_{s}-1}} \nonumber \\
& \times {\frac{(m_{e}-1)!(m_{s}-1)!}{\left[ 
a\frac{(\Omega _{e}\rho _{e}\delta _{e}^{2}+\rho _{e}N_{0}+\sigma 
_{e}^{2})\gamma _{e}}{\rho _{e}\bar{\gamma }_{e}(1-\delta 
_{e}^{2})}-\frac{m_{s}(\Omega _{s}\rho _{s}\delta _{s}^{2}+\rho 
_{s}N_{0}+\sigma _{s}^{2})\gamma _{e}}{\rho _{s}(1-\delta 
_{s}^{2})\bar{\gamma }_{s}}\right] C(a!)^{z}b! }}.
\label{Nak_sp_in_cap}
\end{align}

\section{Numerical results}
This section provides some numerical examples for the derived analytical results, which are also validated by simulations. Unless stated otherwise, the simulation parameters for generation of plots in this section are provided in Table \ref{tabby}.
\begin{table}[hpt]
\centering
\caption{Simulation Parameters.}
\begin{tabular}{|c|l|c|}
\hline
\textbf{S No.} & \textbf{Parameter}          & \textbf{Value} \\ \hline
1.             & Signal processing noise variance $\sigma_s^2=\sigma_e^2$ & 0 dB\\ \hline
2.             & Nakagami-$m$ shape factor $m_s=m_e$    & 2             \\ \hline
3.             & Channel realizations                   & $10^5$                     \\ \hline
4.             & Antenna noise variance $N_0$           & 0.1 dB                    \\ \hline
5.             & Rician $K$ factor $K_s=K_e$                    & 5             \\ \hline
6.             & Main link power $\Omega_s$             & 30 dB              \\ \hline
7.            & Channel estimation accuracy $\delta_s=\delta_e$   & 0.2             \\ \hline
8.            & No. of eavesdroppers $N$               & 5              \\ \hline
9.             & Wiretap link power $\Omega_e$          & 10 dB             \\ \hline
10.             & Power splitting factor $\rho_s=\rho_e$ & 0.8             \\ \hline
\end{tabular}
\label{tabby}
\end{table}

\begin{figure}
  \centering
  \begin{tabular}[!hpt]{c}
    \includegraphics[width=.35\textwidth]{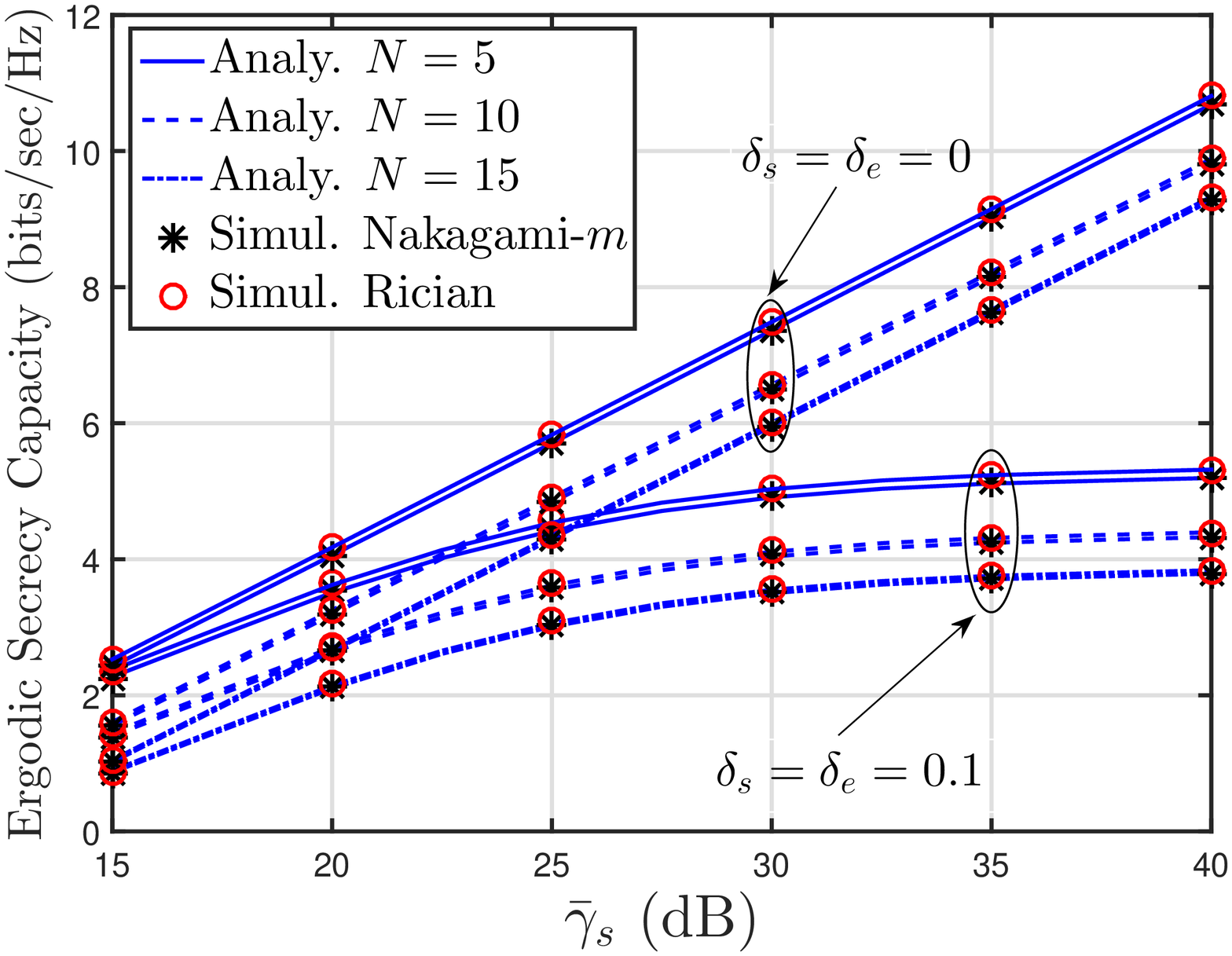}\\
        (a)
  \end{tabular}\qquad
    \begin{tabular}[!hpt]{c}
    \includegraphics[width=.35\textwidth]{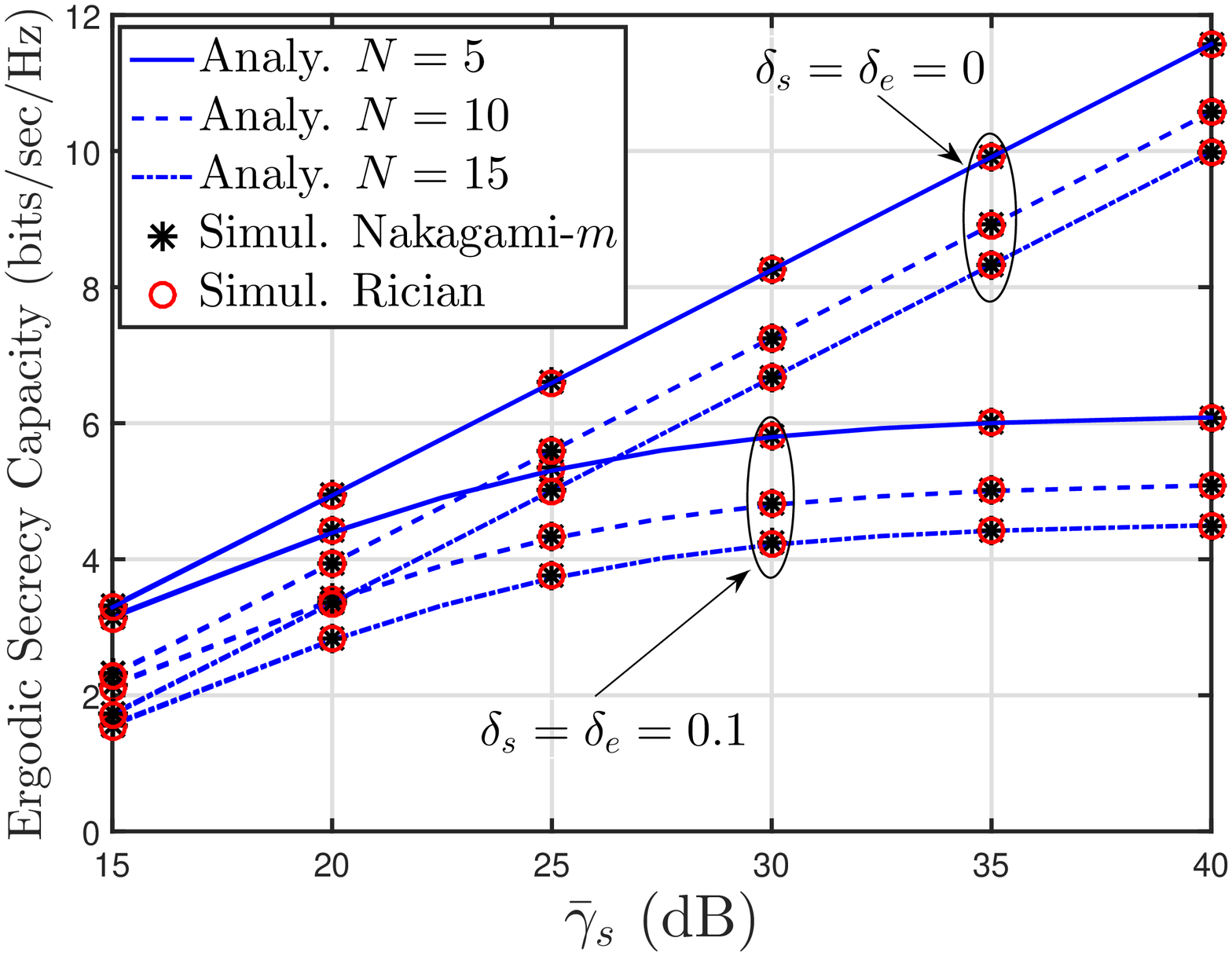}\\
    (b)
        \end{tabular}%
    \caption{Ergodic Secrecy Capacity for Rician and Nakagami-$m$ channels when (a) $S$ and $E$ employ separated design for receivers (b) $S$ employs separated and $E$ use integrated design for receivers.}
    \label{Ric_Nak_2_figs}
\end{figure}

Figure \ref{Ric_Nak_2_figs} (a) plots the ergodic secrecy capacity against increasing $\bar{\gamma}_s$. Note that the ergodic secrecy capacity generally increases with the increase in $\bar{\gamma}_s$. However, the increase in secrecy capacity is hampered due to imperfect channel estimation errors. Moreover, we observe that for particular values of $m_s=m_e$ and $K_s=K_e$, the secrecy capacity for Rician and Nakagami-$m$ channel becomes similar when a large number of eavesdroppers are present in the network. Figure \ref{Ric_Nak_2_figs} (b) further shows the impact receiver design on secrecy performance of energy harvesting IoT devices. The secrecy capacity for Rician and Nakagami-$m$ channel becomes the same despite the large difference in their shape parameters when $\mathcal{S}$ and $\mathcal{E}$ are equipped with separated and integrated receivers. This shows that the impact of wireless channel reduces when $\mathcal{S}$ employs separated and $\mathcal{E}$ employ integrated receiver architecture. 

\begin{figure}[!hpt]
\centering
\includegraphics[width=.35\textwidth]{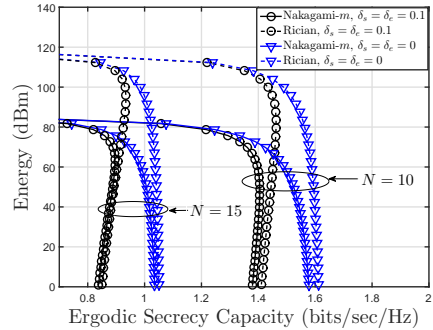}
\caption{Secrecy-energy region with $S$ and $E$ using separated design of receivers, where, $\zeta=0.9$.}
\label{Ric_Nak_4_fig}
\end{figure}

Figure \ref{Ric_Nak_4_fig} plots the secrecy-energy region for Nakagami-$m$ and Rician fading channels. It can be observed that ideal channel estimation always achieves better secrecy-energy pair than that of imperfect channel estimation with $\delta_s=\delta_e=0.1$. Moreover, when $N$ increases from 10 to 15, the secrecy-energy performance significantly decreases. This shows that no. of eavesdroppers have a significant impact on a secrecy-energy pair. Furthermore, the gap between Nakagami-$m$ and Rician channel shrinks for large no. of eavesdroppers.

\section{Conclusion}
In order to meet the future requirements of IoT networks, it is imperative that secure and environment-friendly solutions are proposed. With this intent, we have analyzed an information and power transfer IoT network from the link security perspective. Particularly, we derived closed-form expressions of ergodic secrecy capacity for two small-scale fading models. We also adopted a more practical approach by considering the availability of imperfect channel information at the transmitter. It was shown that the secrecy performance considerably deteriorates under imperfect channel estimation. In other words, an error ceiling is introduced whereby the secrecy capacity cannot be increased more even with the improvements in the SNR of the main link. This trend was observed for different receiver architectures of energy-harvesting IoT devices. Moreover, an increase in the number of eavesdroppers or channel estimation errors causes a significant reduction in the secrecy-energy region for both Rician and Nakagami-$m$ channels. These findings can be considered important from the design perspective of secure RF energy harvesting future networks.

\section*{Acknowledgment}
This work was supported in part by the Academy of Finland under Project No. 311760 and Project No. 319003.

\bibliographystyle{IEEEtran}
\bibliography{References}

\end{document}